\documentclass[%
 reprint,
superscriptaddress,u
 amsmath,amssymb,
 aps,
 prl,
 noeprint,
]{revtex4-2}

\usepackage{graphicx}
\usepackage{subfigure}
\usepackage{physics}
\usepackage{color}
\usepackage{dcolumn}
\usepackage{bm}
\usepackage{comment}
\usepackage{lipsum}
\usepackage{soul}

\newcommand{\etil}[1]{\langle\tilde{#1}\rangle}
\newcommand{\ave}[1]{\langle{#1}\rangle}

\begin{document}

\preprint{APS/123-QED}

\title{Fixing the rotating-wave approximation for strongly-detuned quantum oscillators} 

\author{Jan Ko\v{s}ata}
\thanks{These authors contributed equally to this work.}
\author{Anina Leuch}
\thanks{These authors contributed equally to this work.}
\author{Tobias K\"{a}stli}
\affiliation{Institute for Theoretical Physics, ETH Z{\"u}rich, 8093 Z{\"u}rich, Switzerland}
\author{Oded Zilberberg}
\affiliation{Institute for Theoretical Physics, ETH Z{\"u}rich, 8093 Z{\"u}rich, Switzerland}
\affiliation{Department of Physics, University of Konstanz, 78464 Konstanz, Germany}

\date{\today}

\begin{abstract}
     Periodically-driven oscillators are commonly described in a frame co-rotating with the drive and using the rotating-wave approximation (RWA). This description, however, is known
    to induce errors for off-resonant driving. Here we show that the standard quantum description,
    using creation and annihilation of particles with the oscillator’s natural frequency, necessarily leads to incorrect results when combined with the RWA. We demonstrate this on the simple harmonic oscillator and present an alternative operator basis which reconciles the RWA with off-resonant driving. The approach is also applicable to more complex models, where it accounts for known discrepancies. As an example, we demonstrate the advantage of our scheme on the driven quantum Duffing oscillator.
\end{abstract}

\maketitle

Nature displays many degrees of freedom that are coupled to one another. For simplicity, we describe them as linear resonators that drive each other as they oscillate with a bare harmonic around their respective equilibrium points. Hence, finding the response of a system to an applied drive is an essential task in all areas of science, ranging from fundamental effects in classical and quantum physics to applications in spectroscopy and diagnostics of our daily lives.

The bare harmonic translates to a drive that is periodic in time with a given frequency. To describe the system's response to the drive, a common approach involves moving to a rotating frame, where the system’s behavior appears stationary on short timescales. In the rotating frame, the so-called rotating-wave approximation (RWA) is used, whereby any remaining rapidly oscillating terms are dropped. This procedure becomes less precise when the system and drive frequencies are far detuned from one another, i.e., when the neglected terms are less rapid. Nevertheless, the RWA is a standard approach to find the response of driven systems 
~\cite{Lorch_2018, Lorch_2019}, such
as nanomechanical resonators~\cite{Bachtold_2022, Heugel_2019, Rocheleau_2010}, optical cavities~\cite{Munoz_2021, Rota_2019, Quach_2022, Ferri_2021, Soriente_2021, Soriente_2020, Delpino_2016}, phononic and magnonic modes~\cite{Xu_2021, Delpino_2021, Li_2021, Qi_2021, Gonzalez-Ballestero_2022, Fukami_2021, Marsh_2021} and superconducting junctions~\cite{Blais_2021, Gu_2017, Xiang_2013}. Similarly, it is a common starting point for Floquet engineering and perturbative expansions of nonlinear driven systems~\cite{Mikami2016, Eckardt2017,Eckardt2015,Bukov2015,Goldman2014}.

Despite its ubiquity, the RWA is known to induce errors at large detuning~\cite{Ann_2021}. There the neglected oscillating terms become significant and must be included as perturbations that affect the system's response~\cite{Zheng_2008, Gan_2010, Zhang_2015, Zeuch_2020, Zueco_2009}. Yet, such a perturbative treatment becomes very tasking in many cases that rely on detuned driving, such as optomechanical cooling~\cite{Liu_2013, Marquardt_2008} and the formation of frequency combs~\cite{Weng_2022, Herr_2012, Chembo_2016, Lugiato_1987, Lugiato_2018}. These cases are therefore commonly treated
using the RWA. The issue
is particularly salient in the case of the simple quantum harmonic
oscillator, where the solution obtained with the RWA disagrees
with the exactly soluble classical limit. 

In this work, we pinpoint the source for the quantum-to-classical discrepancy in the driven harmonic oscillator and correct for it. Specifically, we show the description of the oscillator in terms of its bare excitations is ill-suited for the RWA. Instead, we propose a non-standard operator basis (frame) in which the RWA yields the correct solution for the driven case. We furthermore demonstrate that this basis choice is also beneficial for describing driven nonlinear systems, as it significantly improves the fidelity of the RWA in the driven Duffing oscillator at large detuning. We expect that our approach will lead to a significant improvement in the description of numerous scenarios where the response of an oscillator to a coherent drive is calculated.

\begin{figure} 
    \centering
        \includegraphics[width=0.49\textwidth]{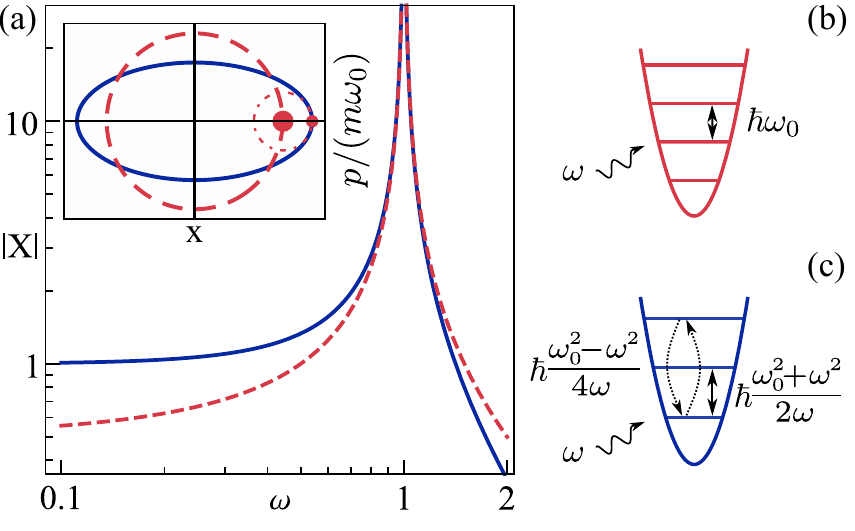}
    \caption{(a) Response amplitude [cf.~Eq.~\eqref{eq:exact_g}] of the driven harmonic oscillator as a function of the driving frequency $\omega$ for the exact solution (solid, blue) and the standard RWA solution (dashed, red) with $F_0=m=\omega_0=1$. The inset shows the corresponding phase space plots for large detuning, $\omega = 0.5$. Spheres indicate solutions in the rotating frame. The ansatz of stationary $\etil{a}$~(large red dashed circle) does not match the exact result. For this, an additional time-dependent term in the solution is needed~(small red circle). This term oscillates at $2\omega$, which results in an elliptical path.  (b) Energy quantization of the harmonic oscillator Hamiltonian~\eqref{eq:rot_H} with energy spacing $\hbar \omega_0$ using the operators $a,\,a^\dagger$ defined in Eq.~\eqref{eq:as}. (c) With the operators $b, b^\dagger$, the harmonic oscillator is not diagonalized, showing a level spacing of $\hbar (\omega_0^2+\omega^2)/ 2\omega$ and processes which create or annihilate two particles with amplitude $\hbar (\omega_0^2-\omega^2)/4\omega$.}
    \label{fig:fig1}
\end{figure}

The Hamiltonian of a periodically driven harmonic oscillator reads
\begin{equation} \label{eq:qho}
H =p^2/2m+m\omega_0^2x^2/2-F_0\cos(\omega t) x\,,
\end{equation}
with position $x$, its conjugate momentum $p$, mass $m$, and resonance frequency $\omega_0$. The driving force has amplitude $F_0$ and oscillates at frequency $\omega$. The time evolution of the system is given by Hamilton's equations of motion (EOMs),
\begin{equation} \label{eq:sho}
\dot{p} = -m \omega_0^2 x + F_0 \cos(\omega t) \,, \quad \dot{x} = p / m \,.
\end{equation}
These EOMs can be solved exactly by transforming to the Fourier domain, where the driving term appears as a Dirac delta function. As a result, the stationary response of the oscillator to the drive reads~\footnote{The effect of initial conditions disappears at sufficiently long times due to dissipation, which we do not consider explicitly here. The results thus correspond to the limit of infinitesimal dissipation and $t\rightarrow \infty$.}
\begin{equation} \label{eq:exact_g}
x = X \cos(\omega t) \,, \quad p = -m\omega X \sin(\omega t)\,,
\end{equation}
with $X =F_0/\left[m(\omega_0^2 - \omega^2)\right]$.
We can parameterize the oscillator's phase space by $x$ and $p/ m\omega_0$. The trajectory described by Eq.~\eqref{eq:exact_g} is then an ellipse with vertices at $X$ and $X \omega/\omega_0$, see Fig.~\ref{fig:fig1}(a).

Within the framework of quantum mechanics, the Hamiltonian~\eqref{eq:qho} is an operator, written in terms of a pair of non-commuting operators $\hat{x}$ and $\hat{p}$.
To find the oscillator’s response in the quantum realm, one commonly introduces raising and lowering operators
$a$ and $a^\dagger$ that diagonalize the time-independent
part of $H$,
\begin{equation} \label{eq:as}
\hat{x} =\sqrt{\frac{\hbar}{2m\omega_0}}(a^\dagger+a)\,, \quad \hat{p} = i\sqrt{\frac{\hbar m \omega_0}{2}}(a^\dagger - a)\,, 
\end{equation}
resulting in
\begin{equation} \label{eq:Hint}
H = \hbar \left[ \omega_0(a^\dagger a+1/2)-  F_a(e^{i\omega t} +e^{-i\omega t})(a^\dagger+a) \right]\,,
\end{equation}
with $F_a=F_0/(2\sqrt{2m\omega_0 \hbar})$, see Fig.~\ref{fig:fig1}(b). The driving renders the Hamiltonian~\eqref{eq:Hint} time-dependent and several methods exist to find the resulting response. For example, one approach involves time-dependent perturbation theory~\cite{Sakurai_1995}, where we move to the interaction picture using the transformation $V(t)\equiv e^{-i a^\dagger a \,\omega_0 t}$, and diagrammatically expand the remaining interaction Hamiltonian $H_{\rm int}$. While this approach is possible for the harmonic oscillator, it rapidly becomes infeasible when additional terms are included in $H$. Hence, in many cases only linear response is considered, e.g., using the input-output method~\cite{Walls_2007}.

Another common approach involves the rotating-wave approximation (RWA). Here, we aim at removing the oscillatory time-dependence altogether. To do so, we first use the unitary transformation $U_a(t) = e^{-i \omega t a^\dagger a}$. In the resulting rotating frame, we obtain the (still time-dependent) effective Hamiltonian
\begin{align}\label{eq:rot_H}
\tilde{H} &\equiv U_a^\dagger H U_a - i \hbar U_a^\dagger \dot{U}_a \\
&=\hbar\left\{-\Delta \tilde{a}^\dagger \tilde{a} - F_a\left[ \tilde{a} \left (1+e^{-2i\omega t} \right) + \tilde{a}^\dagger \left(1 + e^{2i\omega t}\right) \right] \right\} \,,\nonumber
\end{align}
where $\tilde{a} \equiv U_a(t)^\dagger a U_a(t)$ denotes the operator $a$ in the rotating frame, and $\Delta=\omega - \omega_0$ is the so-called detuning away from resonance.
 
In the rotating frame, the remaining time-dependent terms are far-detuned since $2\omega\gg \abs{\Delta}$ and therefore elicit negligible response. Hence, we can apply the RWA and remove all of these from the Hamiltonian~\eqref{eq:rot_H} to obtain $\tilde{H}_{\rm RWA}$. 
We may then find the response by solving Heisenberg's EOM, $i \frac{d}{dt} \etil{a} \equiv  \langle{[\tilde{a}, \tilde{H}_{\rm RWA}]}\rangle / \hbar = -\Delta \etil{a} -F_a$, and obtain the stationary solution $\etil{a} = -F_a / \Delta$. Then, by inverting the transformation $U_a(t)$ we have $\expval{a} = \etil{a} e^{-i \omega t}$. An analogous EOM is similarly solved for $\langle \tilde{a}^\dagger \rangle$. Combining the two solutions [cf.~Eq.~\eqref{eq:as}], we obtain in the classical limit
\begin{equation} \label{eq:x_RWA}
x_\text{RWA}(t) = -\frac{F_0/m}{2\omega_0 \Delta} \cos(\omega t)\,.
\end{equation}

Crucially, the response amplitude found with the RWA does not match the analytical solution in Eq.~\eqref{eq:exact_g}, see Fig.~\ref{fig:fig1}(a). Indeed, the ratio
\begin{equation}
    \frac{x(t)}{x_\text{RWA}(t)} = \frac{2\omega_0\Delta}{\omega^2-\omega_0^2}= \frac{2\omega_0}{\omega_0+\omega}\,,
\end{equation}
only approaches unity for $\Delta \rightarrow 0$, which demonstrates the limitation of the RWA to near-resonant drives. A comparison  between the exact and RWA results in phase space is also illustrative. To this end, we take a stationary amplitude $\etil{a} \equiv \abs{\etil{a}} e^{i \theta}$, and using Eq.~\eqref{eq:as}, we find the equivalent trajectory in phase space, $x(t) = 2\abs{\etil{a}} \sqrt{\frac{\hbar}{2 m \omega_0}} \cos(\omega t - \theta)$ and $p = -2\abs{\etil{a}}\sqrt{\frac{\hbar m \omega_0}{2}} \sin(\omega t-\theta)$. The resulting path is always circular, unlike the elliptical path of the exact solution, cf.~Eq.~\eqref{eq:exact_g} and Fig.~\ref{fig:fig1}(a).
Indeed, the dropped time-dependent oscillating terms, dubbed micromotion, add on to the circular path to produce the full elliptical motion, cf.~small circle in Fig.~1(a). This implies that to describe the elliptical path, $\etil{a}$ must be time-dependent, which cannot be obtained only by time-independent corrections from a high-frequency expansion~\cite{Mikami2016, Eckardt2017,Eckardt2015,Bukov2015,Goldman2014}.  
In other words, the assertion of a stationary $\etil{a}$ violates Hamilton's EOMs, specifically the relation $p = m \dot{x}$. 

As the driven quantum harmonic oscillator is exactly solvable, the RWA was not necessary in the aforementioned analysis. Without the approximation, we would still have obtained the exact result. However, when the system becomes nonlinear, its EOMs are usually not exactly solvable. The RWA, being effectively the lowest order of a perturbative expansion, is then a common method to find the stationary response. As we have seen, however, it neglects crucial corrections due to the detuned drive frequency.
We identify that this systemic mismatch stems from the usual starting point~\eqref{eq:as}, which relies on operators that diagonalize the bare oscillator Hamiltonian and describes excitations with energies $\hbar \omega_0$, see Fig.\,\ref{fig:fig1}(b). 

We now turn to our scheme for an exact treatment of the driven harmonic oscillator. We seek operators akin to $a$ and $a^\dagger$ which recover the ellipsoidal phase-space path corresponding to drive frequency $\omega$. A natural choice is to replace $\omega_0$ by $\omega$ in Eq.~\eqref{eq:as}, i.e., to define new operators $b$ and $b^\dagger$ via
\begin{equation} \label{eq:bs}
\hat{x} =\sqrt{\frac{\hbar}{2m\omega}}(b^\dagger+b)\,, \quad \hat{p} = i\sqrt{\frac{\hbar m \omega}{2}}(b^\dagger - b)\,,
\end{equation}
which satisfy $[b, b^\dagger]=1$, and describe the system in terms of excitations with the drive's frequency $\omega$. The driven harmonic oscillator Hamiltonian~\eqref{eq:qho} now reads
\begin{align} 
H = \hbar \bigg\{ &\frac{\omega_0^2 + \omega^2}{2 \omega}  \left (b^\dagger b + \frac{1}{2} \right)+ \frac{\omega_0^2  - \omega^2}{4 \omega} \big(b^2 + (b^\dagger)^2  \big)\nonumber
 \\ &- F_b(e^{i\omega t} +e^{-i\omega t})(b^\dagger+b) \bigg\} \,,\label{eq:Hb}
\end{align}
with $F_b = F_0 / (2 \sqrt{2 m \omega \hbar})$, see Fig.~\ref{fig:fig1}(c). Note that we effectively rotated from the $a$ to the $b$ operators using a unitary (Bogoliubov) transformation~\cite{Xiao_2009}. With the new operators, Eq.~\eqref{eq:Hb} contains squeezing terms that do not conserve the particle number, since $b$ and $b^\dagger$ do not diagonalize the bare harmonic oscillator unless $\omega = \omega_0$. 

We repeat the same procedure in the new basis. We first apply the unitary transformation $U_b(t) = e^{-i \omega t b^\dagger b}$ to move to a rotating frame, and then obtain the corresponding Heisenberg's EOMs
\begin{equation} \label{eq:Heis_corr}
i \frac{d}{dt} \etil{b} = \frac{\omega_0^2 - \omega^2}{2\omega} \left( \etil{b} +  \langle \tilde{b}^\dagger \rangle e^{2 i \omega t}\right) - F_b\left(1  + e^{2 i \omega t} \right)\,.
\end{equation}
Crucially, when we now search for a stationary solution for $\etil{b}$, we obtain
\begin{equation}
    \etil{b} = \frac{2F_b\,\omega}{\omega_0^2 - \omega^2} \,,
    \label{eq:main_res}
\end{equation}
which upon inverting $U_b(t)$ matches the exact result in Eq.~\eqref{eq:exact_g}. Importantly, the correct result is obtained regardless of whether we take the RWA or not. In other words, when the operators $b, b^\dagger$ are used, the stationary solution in the rotating frame is exact and satisfies Eq.~\eqref{eq:Heis_corr}. This is the main result of this work.

\textit{Example: Duffing oscillator.---}
The RWA is a common starting point for dealing with time-dependent systems that are not exactly solvable. Our result~\eqref{eq:main_res} implies that a suitable choice of operators made prior to applying the RWA can significantly reduce errors at large detuning. To demonstrate this, we compare the results obtained using the RWA with the two operator definitions, $a,a^\dagger$ and $b,b^\dagger$. We consider a Duffing oscillator, described by the Hamiltonian
\begin{equation} \label{eq:H_Duffing}
H_D = p^2/2m + m \omega_0^2 x^2/2 + \alpha x^4/4 - F_0 \cos(\omega t)  x\,,
\end{equation}
where $\alpha$ is the Duffing nonlinearity. We plug both operator definitions [Eqs.~\eqref{eq:as} and~\eqref{eq:bs}] into Eq.~\eqref{eq:H_Duffing}, move to a rotating frame [cf.~Eq.~\eqref{eq:rot_H}], and apply the RWA in both procedures. Note that the RWA drops multiple oscillating terms that describe frequency conversion processes due to the Duffing nonlinearity. We furthermore apply a semiclassical mean-field ansatz~\footnote{This allows us to take $\langle \tilde{a}^3 \rangle \rightarrow \etil{a}^3$ etc.} to obtain Heisenberg's EOMs
\begin{align} \label{eq:duff_a}
    i \frac{d}{dt} \etil{a} &=
    -\Delta \etil{a}-\frac{F_a}{\hbar}+\frac{3 \alpha \hbar}{4m^2\omega_0^2}\left(\etil{a}+\etil{a}^2\ave{\tilde{a}^\dagger}\right)\,,\\
\label{eq:duff_b}
    i \frac{d}{dt} \etil{b} &= 
    \frac{\omega_0^2 - \omega^2}{2\omega}\etil{b}-\frac{F_b}{\hbar}+\frac{3\alpha \hbar}{4m^2\omega^2}\left(\etil{b}+\etil{b}^2\ave{\tilde{b}^\dagger}\right)\,.
\end{align}

\begin{figure}
    \centering
        \includegraphics[width=0.49\textwidth]{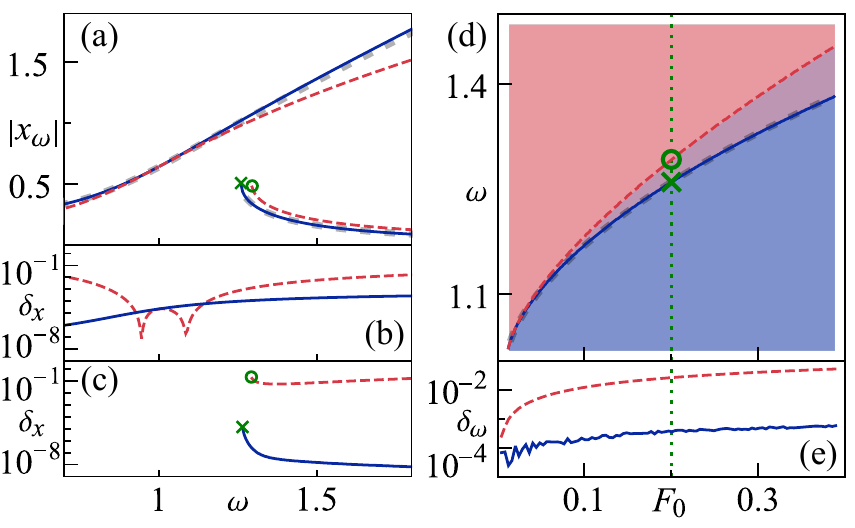}
    \caption{A comparison of the Duffing oscillator response obtained using the $a, a^\dagger$ and $b, b^\dagger$ operator bases [cf.~Eqs~\eqref{eq:as},~\eqref{eq:bs}] with a time-dependent simulation. The parameters used are $m=\omega_0=\alpha=1$. (a) The response amplitude $\abs{x_\omega}$ plotted against the drive frequency $\omega$ for $F_0=0.2$ using $a, a^\dagger$ (dashed, red) and $b,b^\dagger$ (solid, blue). The time-dependent simulation result (dot-dashed grey, behind blue) consists of adiabatic up and downsweeps of $\omega$.
    (b), (c) The relative discrepancy in log scale of the $a,a^\dagger$ (dashed, red) and the $b, b^\dagger$ (solid, blue) results from the time-dependent simulation. 
    The circles and crosses in (a) and (c) represent the jump frequency in the downsweep for operators $a$ and $b$.
    (d) A phase diagram as a function of $\omega$ and $F_0$. Blue (red) fill denotes regions with one (two) stable solution(s). 
    (e) Relative discrepancy $\delta_{\omega} = \abs{(\omega - \omega_{\text{RWA}}) \,/\, \omega }$ in log scale of the phase boundaries obtained using $a, a^\dagger$ (dashed, red) and $b, b^\dagger$ (solid, blue) from the time-dependent result. The numerical approach is exact up to $10^{-5}$ which becomes visible at low $F_0$, where the discrepancy of our scheme is of that order. The green dotted lines in (d) and (e) indicate where the simulation of (a) takes place.
    }
    \label{fig:fig2}
\end{figure}

We can now search for stationary solutions in either of the rotating frames [$\frac{d}{dt} \etil{a} = 0$ or $\frac{d}{dt} \etil{b} = 0$]. Both Eqs.~\eqref{eq:duff_a} and~\eqref{eq:duff_b} generate a cubic polynomial condition which has up to three solutions~\cite{Kosata_2022, Heugel_2019, Lifshitz_2008}, see Fig.~\ref{fig:fig2}(a). Both approaches produce the expected tail-shaped response, i.e., a single solution in the $\omega < \omega_0$ regime that bends up towards a high-amplitude solution and a coexistence region in the $\omega > \omega_0$ regime, where both low- and high-amplitude solutions appear. 
The coexistence region manifests mathematically as a bifurcation point for the roots of the polynomial condition. Despite qualitative agreement, the solutions in the two frames differ quantitatively, i.e., in their amplitudes as a function of detuning and in the positions of their bifurcation points. 

To find out which of the two approximate procedures describes the driven Duffing oscillator~\eqref{eq:H_Duffing} more accurately, we compare their predictions with a ``numerical experiment'' in the classical picture. Specifically, we numerically evolve Hamilton's equations for the Hamiltonian~\eqref{eq:H_Duffing} until we reach stationary motion for different detunings and initial conditions and then adiabatically sweep the detuning to explore the solution landscape. Note that we add an infinitesimal dissipation term to enforce convergence in the simulation, which negligibly shifts the stationary outcome. Additionally, the numerical time-trace describes the oscillatory response at the drive frequency $\omega$ including high harmonic generation. Hence, the relevant benchmark for the RWA solutions is the Fourier component $x_\omega$ at frequency $\omega$ of the time trace. In Figs.~\ref{fig:fig2}(b) and (c), we plot the relative discrepancy $\delta_x = \abs{(x_\omega - x_{\omega, \text{RWA}})\, /\, x_\omega}$ between the numerically obtained $x_\omega$ and the two RWA results in the high- and low-amplitude solution branches, respectively. We observe that our new RWA approach agrees much better for all values of $\omega$. In the low-amplitude regimes, the system responds quasi-linearly and our scheme performs better as discussed above in the harmonic oscillator case. Slight deviations are seen in the high-amplitude regime, where nonlinear effects dominate the physics.

In driven-dissipative systems, phase diagrams depict the number or type of stationary solutions as a function of system parameters~\cite{Soriente_2018, Chitra_2015}. Here, we focus on the location of the bifurcation point as a function of detuning and drive amplitude.
The bifurcation point is marked in Figs.~\ref{fig:fig2}(a) and (c), where the better performance of our RWA approach manifests as a clear change in its position. In Fig.~\ref{fig:fig2}(d), we draw phase diagrams obtained by the two RWA approaches, and observe a distinct difference in the phase boundaries. The exact numerical solution matches our scheme much better, see Fig.~\ref{fig:fig2}(e).

\textit{Conclusion.---}
In summary, we have presented an operator basis for periodically-driven systems which anticipates a response at the driving frequency and is thus better suited for using the RWA. For both the harmonic and the Duffing oscillators, this basis choice significantly improves the fidelity of the RWA while retaining its simplicity. Our approach is applicable to systems in many areas of physics beyond mechanical oscillators described by $x$ and $p$~\cite{Bachtold_2022}. In quantum optics~\cite{Walls_2007} and optomechanics~\cite{Aspelmeyer_2014}, a driven cavity mode is described by Eq.~\eqref{eq:qho} with the magnetic and electric fields playing the roles of $x$ and $p$. For models with inherent detuning such as optomechanical cooling and frequency-comb generation, our approach may result in significant numerical corrections. A specific case of interest are models coupling a cavity to one or more two-level systems~\cite{Shore_1993,Kirton_2019}; here the so-called general RWA is used, where a basis diagonalizing the spin-cavity interaction reduces the RWA error~\cite{Irish_2007}. As general RWA uses the standard operators $a$ and $a^\dagger$, we expect that it can be improved further by using our scheme. Detuned driven systems also appear in circuit QED where the magnetic flux and charge appear in place of $x$ and $p$~\cite{Blais_2021}. Overall, a sizable body of work exists elaborating on the errors due to the RWA at large detuning~\cite{Baker_2018, Niemczyk_2010, Bishop_2010, Sornborger_2004}. Our work suggests that in many cases, these errors can be dramatically reduced by choosing an appropriate operator basis.

\begin{acknowledgments}
The authors would like to thank T. L. Heugel, J. del Pino, A. Grimm, A. Eichler and G. Blatter for valuable discussions. This work was supported by the Swiss National Science Foundation through grant CRSII5 1771981. O.Z. acknowledges funding through SNSF grant PP00P2\_190078 and from the Deutsche Forschungsgemeinschaft (DFG) - project number 449653034.
\end{acknowledgments}

\bibliography{bibliography.bib}

\end{document}